\documentclass[aps,twocolumn,eqsecnum,bibnotes,showpacs]{revtex4}
\usepackage{graphicx}
\usepackage{bm}
\usepackage[tbtags]{amsmath}

\def\sfrac#1#2{{\ts\frac{#1}{#2}}}
\def\beq{\begin{equation}}
\def\eeq{\end{equation}}
\def\eeql#1{\label{#1} \end{equation}}
\def\f{\bm{f}}

\def\ot{\! \otimes \!}
\def\om{\omega}
\def\H{{\cal H}}
\def\bdot{{\mbox{\boldmath $\cdot$}}}

\def\sgn{\mathop{\rm sgn}\nolimits}
\def\diag{\mathop{\rm diag}\nolimits}
\def\ts{\textstyle}

\makeatletter
\def\NESW{\mathinner {\mkern 1mu\raise \p@ \hbox {.}\mkern 2mu \raise 4\p@ \hbox {.}\mkern 2mu\raise 7\p@ \vbox {\kern 7\p@ \hbox {.}}\mkern 1mu}}
\makeatother
\begin{document}

%==========================================================

\title{Hamiltonian and Linear-Space Structure for Damped Oscillators: II. Critical Points}

\author{S.C. Chee}
\author{Alec \surname{Maassen van den Brink}}
\thanks{Corresponding author;\\ electronic address: \texttt{alec@dwavesys.com}}
\author{K.~Young}
\affiliation{Physics Department, The Chinese University of Hong Kong, Hong Kong, China}

\date{first posted on 17 Jun 2002; revised on 10 Feb 2004}

%==========================================================

\begin{abstract}
The eigenvector expansion developed in the preceding paper for a system of damped linear oscillators is extended to critical points, where eigenvectors merge and the time-evolution operator $\H$ assumes a Jordan-block structure.  The representation
of the bilinear map is obtained in this basis. Perturbations $\epsilon \Delta \H$
around an $M$-th order critical point generically lead to eigenvalue shifts $\sim \epsilon^{1/M}$ dependent on only \emph{one} matrix element, with the $M$ eigenvalues splitting in equiangular directions in the complex plane.  Small denominators near criticality are shown to cancel.
\end{abstract}


\maketitle

%==========================================================

\section{Introduction}
\label{sect:intro}

The preceding paper~\cite{pap1} [hereafter referred to as I and equations therein as e.g.\ (I.2.3)] developed an eigenvector expansion for a broad class of systems with $N$ ohmically damped coupled oscillators, also applicable to interacting quantum systems. The key concept is a bilinear map $(\bm{\psi},\bm{\phi})$, under which the time-evolution operator $\H$ is symmetric: $(\bm{\psi},\H\bm{\phi})=(\H\bm{\psi},\bm{\phi})$, thus allowing concepts familiar from conservative systems to be transcribed.

In I, it is assumed that (a)~the eigenvectors $\f_j$ of $\H$ are complete and (b)~$(\f_j,\f_j)\ne0$ for all $j$. These are violated (together) only at critical points (with measure zero in parameter space), where eigenvectors merge. This case is rather more technical, and $\H$ takes on a Jordan-block (JB) structure. Some results are known in a continuum limit~\cite{jbsr}, but this paper gives a self-contained account, putting the concept of critical damping into a general framework, in which projections etc.\ can still be performed using the bilinear map.

Since merged eigenvectors no longer span the whole space, a basis has to be constructed---the well-known Jordan normal basis~\cite{jb0}. Here, we need to represent the bilinear map (viz., the metric $g_{\bdot\bdot}$) as well and verify that the resulting $\H_{\bdot\bdot}$ is symmetric (Section~\ref{sect:jb}). Then time evolution (Section~\ref{sect:time}) and perturbation theory (Section~\ref{sect:jbpert}) are developed in this basis. The latter also permits a discussion of small denominators near criticality.

An $N=1$ example (Section~III~A in I) already illustrates criticality, and also explains its name. Section~\ref{sect:ex} presents higher-order examples, including perturbations around them. We end with a discussion in Section~\ref{sect:disc}. An extensive account can be found in Ref.~\cite{chee}.

We stress that when eigen\emph{values} merge (degeneracy), either the eigenvectors merge as well (criticality), or they could remain distinct (level crossing)---though the latter is exceptional in a sense to be made precise in Section~\ref{sect:ex}.

%==========================================================

\section{Jordan blocks}
\label{sect:jb}

For any finite matrix $\H$, a Jordan normal basis is known to exist in general~\cite{jb0}. In a revival of interest in non-hermitian and non-diagonalizable systems~\cite{non-herm}, JBs increasingly have the attention of the physics community. We briefly recapitulate some standard results in linear algebra, and address the remaining fruitful object of study: the \emph{consequences} of JBs for $\H$ \emph{of the specific form} (I.2.3), corresponding to the well-motivated dynamics (I.1.1) and leading to the bilinear map (I.2.9).

\subsection{Basis vectors}
\label{subsect:constr}

Suppose there are $\nu$ independent eigenvectors $\f_j$ with eigenvalues $\om_j$. If $\nu<2N$, the dimensionality of phase space, the $\f_j$ can be augmented into the Jordan normal basis $\{\f_{j,n}\}_{1\le j\le\nu,0\le n\le M_j}$, obeying
\beq
  ( \H - \om_j ) \, \f_{j,n} = \f_{j,n{-}1}\;,
\eeql{eq:x04}
with $\f_{j,-1}\equiv0$ so that $\f_{j,0}=\f_j$. Separating positions and momenta, $\f_{j,n}=(f_{j,n},\hat{f}_{j,n})^\mathrm{T}$. The span of $\{\f_{j,n}\}_{n=0}^{M_j-1}$ for fixed $j$ is called the JB at $\om_j$, with size~$M_j$; in it, $\H$~has the \emph{Jordan normal form}
\beq
 \bar{\H}^{\bdot}{}_{\bdot} = \mbox{\small $
 \begin{pmatrix} \om_j & 1 & \cdots & 0 & 0 \\
     0 & \om_j & \ddots & 0 & 0 \\
     \vdots & \vdots & \ddots &  \ddots & \vdots \\
     0 & 0 & \cdots & \vphantom{\ddots}\om_j & 1 \\
     0 & 0 & \cdots & \vphantom{\ddots}0 & \om_j \end{pmatrix}$}
\eeql{eq:jbh}
with respect to the normal basis. The block structure $\{(\om_j,M_j)\}$ is completely specified by $\H$, but vectors from blocks with identical eigenvalues may be mixed, as already shown by degeneracies in the diagonalizable case. Until Section~\ref{subsect:degen}, however, all $\om_j$ are assumed distinct.

Then, the only arbitrariness consists of
\beq
  \f_{j,n} \mapsto \sum_{k=0}^{n} c_k \f_{j,n{-}k}
\eeql{eq:arb1}
with $c_0\neq0$, leaving (\ref{eq:jbh}) invariant and generalizing $\f_j\mapsto c\f_j$ for simple~$\om_j$. This freedom will be exploited below.

When JBs occur, $J(\om) = \det ( \H^{\bdot}{}_{\bdot} -\om )$ has a multiple zero. But this is \emph{not} sufficient, since it could also correspond to level crossing without JBs. This is why the formalism is developed without reference to~$J$---in contrast to the case of continuum models (Section VI in~I). 

\subsection{Bilinear map}
\label{subsect:jborth}

Although the Jordan normal form itself is standard, we need to consider the bilinear map $(\f_{j,n},\f_{j',n'})$. Using (\ref{eq:x04}) on $(\H\f_{j,n},\f_{j',n'}) =(\f_{j,n},\H\f_{j',n'})$ yields
\begin{multline}
  \om_j ( \f_{j,n} , \f_{j', n'} )+ ( \f_{j,n{-}1} , \f_{j', n'} ) \\
  =\om_{j'} ( \f_{j,n} , \f_{j', n'} )+( \f_{j,n},\f_{j', n'{-}1})\;,\label{eq:orth02} 
\end{multline}
for $n \le M_j-1$ and $n' \le M_{j'}-1$.

Orthogonality for $j \ne j'$ is proved by induction with respect to $n+n'$.  The case $n=n'=0$ is the standard one as in (I.2.11). On each side of (\ref{eq:orth02}), the second term vanishes by the induction hypothesis, leaving 
$(\om_j-\om_{j'})(\f_{j,n},\f_{j',n'})=0$, and completing the induction.

Next consider $j=j'$; with $M\equiv M_j$ and $(n,n')\equiv(\f_{j,n},\f_{j,n'})$, (\ref{eq:orth02}) gives $(n{-}1 , n') = (n, n'{-}1)$, leading to
\beq
  (n, n') = A_{n{+}n'}\;.
\eeql{eq:orth04}
For $n\le M{-}2$, we have $A_n=(-1,n{+}1)=\nobreak0$ [again associating incomplete eigenvectors with $(\f_j,\f_j)=0$]. If $A_{M{-}1}$ would also vanish, then $\f_{j,0}$ would be orthogonal to \emph{every} basis vector, which is impossible [cf.~below (I.2.16)]. We can choose $A_{M{-}1}=\nobreak1$ and $A_n=0$ for $M\le n\le2M{-}2$. To show this, first perform a transform (\ref{eq:arb1}) with $c_k=c_0\delta_{k0}$, under which $A_{M{-}1}\mapsto c_0^2 A_{M{-}1}=1$ for some $c_0\neq0$. Further transforms $c_k=\delta_{k0}+c_n\delta_{kn}$ for $n= 1, \ldots, M{-}1$ take $A_{M{+}n{-}1}=(n, M{-}1 )\mapsto A_{M{+}n{-}1} + 2c_n=0$ by a choice of $c_n$, while $A_{M{+}m{-}1}$ with $m<n$ (taken care of in previous steps) are not affected (neither, of course, are $A_n=0$ for $n<M{-}1$).

All these orthogonality relations are captured by
\beq
  ( \f_{j,n},\f_{j',n'} ) = \delta_{jj'} \delta_{n+n', M_j{-}1}\;,
\eeql{eq:summ1}
and completeness can be written as
\beq
  \bm{\phi}=\sum_{j,n}\f_{j,n}\,(\f_{j,M_j{-}1{-}n},\bm{\phi})\;.
\eeql{eq:comp4}

Equation (\ref{eq:summ1}) can also be stated as a representation for $\bar{g}_{\bdot\bdot}$ in (one block of) the Jordan normal basis:
\beq
  \bar{g}_{\bdot\bdot} = \mbox{\small $
  \begin{pmatrix}0 & 0 & \cdots & 0 & 1 \\
     0 & 0  & \cdots & \vphantom{\ddots} 1 & 0 \\
     \vdots & \vdots & \NESW & \vdots & \vdots \\
     0 & 1 & \cdots & \vphantom{\ddots} 0  & 0 \\
     1 & 0 & \cdots & \vphantom{\ddots} 0 & 0  \end{pmatrix}$}\;.
\eeql{eq:jbflip}
It is now straightforward to verify that $\bar{\H}_{\bdot\bdot}$ is 
\beq
 \bar{\H}_{\bdot\bdot} = \mbox{\small $
  \begin{pmatrix} 0 & \vphantom{\ddots}0 & \cdots & 0 & \om_j \\
     0 & \vphantom{\ddots}0 & \cdots & \om_j & 1\\
     \vdots & \vdots & \NESW & \NESW & \vdots \\
     0 & \om_j & \NESW &  0 & 0 \\
     \om_j & 1 & \cdots & \vphantom{\ddots}0 & 0  \end{pmatrix}$}\;,
\eeql{eq:jbh2}
a \emph{symmetric} Jordan-type matrix. 

Therefore, the duals defined by [cf.~(I.2.19)]
\beq
  \f^{j,n}\equiv{\cal D}\f_{j,n}=\left[ g_{\bdot\bdot}\f_{j,M_j{-}1-n}\right]^*
\eeql{eq:jbdual}
ensure 
\begin{gather}
  \langle \f^{j,n} | \f_{j',n'} \rangle=\delta^j{}_{j'} \, \delta^n{}_{n'}\;,
  \label{eq:orthjb}\\
  \mathcal{I}=\sum_{j,n}\f_{j,n}\,\langle\f^{j,n}|\,\bdot\,\rangle\;,\label{eq:compjb2}
\end{gather}
the latter in an obvious shorthand. In this case, $\mathcal{D}\bm{\phi}$ cannot be calculated without first resolving it in terms of the basis; contrast the non-critical case in I. Of course, the (unique) dual to any basis $\{\bm{e}_k\}$ can always be calculated by inverting the $2N\times2N$ Gramm matrix $\langle\bm{e}_k|\bm{e}_l\rangle$~\cite{jb0} without any reference to the bilinear map, but this matrix is not sparse for $\{\bm{e}_k\}=\{\f_{j,n}\}$. The above has bypassed the general inversion in that all manipulations below (\ref{eq:orth04}) take place \emph{within} a (typically small) block only, made possible by (\ref{eq:orth02}). Cf.\ the remarks in I, Section~II~E.

Together with our normalization, (\ref{eq:x04}) implies that the counterpart to (I.2.14) for conjugate blocks reads
\beq
  \f_{-j,n}=\pm i^{M_j}(-)^n\f_{j,n}^*\;,
\eeql{JBconj}
where the overall sign should be the same within each block. For a block with imaginary frequency (a so-called zero-mode, see I) which does not cross with another block, we can set $j=\nobreak0$, and (\ref{JBconj}) becomes a symmetry of the basis vectors. The treatment of level crossing follows the one in I and will not be repeated.

Incidentally, in most cases only \emph{one} nontrivial JB (say~$j$) is formed at a given parameter value. Then, the $\f_{j,n}$ can be constructed without having to solve the linear system (\ref{eq:x04}). First, determine~$\f_j$; now $\bm{\psi} = [g_{\bdot\bdot} \f_j]^*$ is guaranteed to have $(\bm{\psi},\f_j)\neq0$. Orthogonalize $\bm{\psi}$ with respect to all the other eigenvectors [which does not change $ (\bm{\psi} , \f_j ) \ne 0$]; the result is $\f_{j,M_j{-}1}$. After the $\f_{j,n}$ with lower $n$ are obtained from (\ref{eq:x04}), the normal basis at $\om_j$ can be biorthogonalized as above.

\subsection{Example}
\label{subsect:excrit}

Consider a single oscillator as in I, Section~III A, but at the critical $k = k_* = \gamma^2$. The eigenvector is $\f_{,0} = c_0(1,-\gamma)^\mathrm{T}$. Then $\f_{,1} = c_0(-i/\gamma,0)^\mathrm{T}+c_1(1,-\gamma)^\mathrm{T}$.

One verifies $A_0=0$ and $A_1=\nobreak c_0^2$, independent of $c_1$; we take $c_0 = 1$. Finally $A_2=2(c_1-i/\gamma)=0$ if $c_1=i/\gamma$.  With these choices, $\f_{,0}=(1,-\gamma)^\mathrm{T}$ and $\f_{,1}=(0,-i)^\mathrm{T}$.

\subsection{Level crossings}
\label{subsect:degen}

When levels cross in conservative systems, orthogonality is only a matter of choice. Here the situation is more subtle still, since blocks of arbitrary size may cross. Suppose that the normal basis has $\om_1 = \om_2 = \ldots = \om_L$~\cite{LleN}. The corresponding vectors [their number given by the order of the zero in $J(\om)$] form an array, e.g.,
\beq
  \begin{array}{cccc}
  \f_{1,3} & \f_{2,3} & & \\
  \f_{1,2} & \f_{2,2} & \f_{3,2} & \\
  \f_{1,1} & \f_{2,1} & \f_{3,1} & \\
  \f_{1,0} & \f_{2,0} & \f_{3,0} & \f_{4,0}
  \end{array}\;;
\eeql{eq:exarray1}
each of the $L$ columns corresponds to a JB, obeying (\ref{eq:x04}). While degeneracy is doubly exceptional (as will be seen in examples), we give the rigorous treatment for the record.

We need to ensure 
\beq
  (\f_{j,n} , \f_{j'n'}) = \delta_{jj'} \delta_{n{+}n',M_j-1}\;.
\eeql{eq:orth}
It is immediately shown, in the usual way, that
\beq
  (\f_{j,n} , \f_{j',n'}) = A^{j,j'}_{n{+}n'}\;,
\eeql{eq:diag}
reducing the number of conditions to be checked. To proceed, we use the freedom
\beq
  \f_{j,n} \mapsto \sum_{j'=1}^L \sum_{k=m}^n c^{j,j'}_k \f_{j',n-k}\;,
\eeql{eq:free}
where $m=\max(0,M_j{-}M_{j'})$. The mixing between blocks is the new feature here, as is apparent from the case of two degenerate \emph{trivial} blocks. The lower limit $k=m$ signifies that large blocks can always mix into smaller ones, but not vice versa. For instance, in (\ref{eq:exarray1}), the only freedom in $\f_{1,0}$ and $\f_{2,0}$ consists in mixing these two vectors among each other.

We start by finding a normalizable ``top" vector. If $M_2<M_1$, then $\f_{1,M_1-1}$ can be normalized by the argument below (\ref{eq:orth04}). If there is no single largest block, consider the quadratic form $\bm{(\phi},(\H{-}\om_j)^{M_1-1}\bm{\phi)}$ in the span of all vectors $\f_{j,M_1-1}$ [in the example (\ref{eq:exarray1}), $j=1,2$]. This form cannot vanish identically, for else the corresponding eigenvectors $\f_{j,0}$ would be orthogonal to the entire space by (\ref{eq:diag}). Take one vector on which the form is nonzero, relabel it as $\f_{1,M_1-1}$, and redefine the other $\f_{j,M_1-1}$ if necessary so that they still span the same subspace. Subsequently, construct the associated lower $\f_{j,n}$ by iterating $\H-\om_j$. Clearly, the basis transformation of this paragraph is of the form (\ref{eq:free}).

In the block $j=1$, one now has precisely the situation of Section~\ref{subsect:jborth}; consequently, this block can be biorthogonalized as described there. It remains to orthogonalize the other blocks with respect to this first one. Then relation (\ref{eq:orth}) will hold for $j=1$ and any $j'$, which will not change when the blocks $j\ge2$ are mixed among themselves upon repeating the whole procedure.

For orthogonalizing the blocks $j\ge2$ with respect to the $j=1$ block, completing one iteration of the construction, it suffices to consider the top vectors $\f_{j,M_j-1}$: when their associated block is reconstructed by iterating $\H-\om_j$, all vectors in it will be orthogonal to $\{\f_{1,n}\}$ by virtue of (\ref{eq:diag}). By the same token, $(\f_{j,M_j-1},\f_{1,n})=0$ already if $n<M_1-M_j$. For $n\ge M_1-M_j$, this bilinear map can be made to vanish by $\f_{j,M_j-1}\mapsto\f_{j,M_j-1}+c^{j,1}_{M_j+n-M_1}\f_{1,M_1-1-n}$. Thus, the needed mixings are precisely those which conserve the block structure and are allowed by (\ref{eq:free}); e.g., in (\ref{eq:exarray1}), in the course of the above we do not mix $\f_{1,3}$ into $\f_{3,2}$.

Iterating the procedure until the number of blocks~$L$ is exhausted, we finally achieve (\ref{eq:orth}) in general. Clearly, all of the above goes through as well if level crossing occurs at several frequencies.

\subsection{Sum rules}
\label{subsect:sum}

Separating (\ref{eq:comp4}) into coordinates and momenta leads to four sum rules, which can be written in terms of the coordinates alone. Using the shorthand $n' = M_j-1-n$, 
\begin{subequations}
\begin{align}
  0 &= \sum_{j,n}f_{j,n} \ot f_{j,n'}\label{sr1} \\
  I &=\sum_{j,n} \left[ \om_j f_{j,n} + f_{j,n{-}1} \right] \ot f_{j,n'}
    \label{eq:sr2} \\
  0 &=\sum_{j,n} \left[ 
  (\om_j^2 f_{j,n} + 2 \om_j f_{j,n{-}1} + f_{j,n{-}2}) \ot f_{j,n'} \right.\notag\\
  &\hphantom{=\sum_{j,n} \bigl[} \left. {} + i(\om_j f_{j,n} + f_{j,n{-}1}) \ot
   (\Gamma f_{j,n'}) \right]\label{sr3} \\
  0 &= \sum_{j,n} f_{j,n} \ot (\Gamma f_{j,n'})\label{sr4}\;.
\end{align}
\end{subequations}%
Here, $a\otimes b$ is the matrix with elements $a(\alpha) b(\beta)$. We have verified these in, e.g., the examples of Section~\ref{sect:ex}.

\subsection{Time Evolution}
\label{sect:time}

Since $\H$ is not diagonal, the basis vectors' time dependence is slightly complicated. The defining $i\partial_t \f_{j,n}(t) = \om_j\f_{j,n}(t) + \f_{j,n{-}1}(t)$ [$\f_{j,n}(t{=}0)\equiv\f_{j,n}$] is solved by
\begin{align}
  \f_{j,n}(t) &=\sum_{l=0}^n C_l(\om_j,t) \f_{j,n-l}\;,\label{eq:time06}\\
  C_l(\om_j,t) &=\frac{ (-it)^l}{l!} e^{-i\om_j t}
  = \frac{1}{l!} [\partial_{\om}^l  e^{-i\om t} ]_{\om=\om_j}\;.\label{eq:time07}
\end{align}
The evolution of a general initial state $\bm{\phi}(t{=}0)\equiv\bm{\phi}$ follows by simply putting $\f_{j,n}\mapsto\f_{j,n}(t)$ in (\ref{eq:comp4}). More formally, the retarded Green's function thus reads
\beq
  \mathcal{G}(t)=\theta(t)\sum_j\sum_{n=0}^{M_j-1}\f_{j,n}(t)
  \langle\f^{j,n}|\bdot\rangle\;.
\eeql{eq:jbgr07}
Since $\theta(t)C_l(\om_j,t)\mapsto i(\om-\om_j)^{-l-1}$ under Fourier transform, this leads to
\beq\begin{split}
  \tilde{\mathcal{G}}(\om)=\sum_j\sum_{n=0}^{M_j{-}1}\sum_{l=0}^n
  \f_{j,n-l}\frac{i}{(\om-\om_j)^{l+1}}\langle\f^{j,n}|\bdot\rangle\;,
\end{split}\eeql{eq:jbgr06}
readily verified to solve $(\H-\om)\tilde{\mathcal{G}}(\om)=-i\mathcal{I}$.

%==========================================================

\section{Jordan-block perturbation theory}
\label{sect:jbpert}

\subsection{Lowest order}
\label{sect:jbpertlow}

Although perturbation theory has been given in I, the situation at a critical point is different, exhibiting interesting features not found in conservative systems.  For simplicity consider only the case without level crossing. Under a perturbation of a critical $\H_0$ given by (\ref{eq:jbh}) in the Jordan normal basis,
\beq
  \H = \H_0 + \epsilon \Delta \H\;,
\eeql{eq:jbp01}
a JB of size $M_j$ to leading order generically behaves as follows. (a)~The eigenvalue splits into $M_j$ different ones, shifting in equiangular directions in the complex frequency-plane and at the same rate.  The directions for $\epsilon > 0$ bisect those for $\epsilon < 0$. (b)~The frequency shifts go not as $\epsilon$ but as $\epsilon^{1/M_j}$. (c)~The shifts depend on only one element of the $M_j \times M_j$ matrix $\Delta \H$ within this block.

These features are already seen in the trivial $N=1$, $M_j=2$ example of I, Section III~A. For $k = k_* + \epsilon$, $\om_j=-i\gamma \pm \sqrt{\epsilon}$, approaching the critical point along the real direction for $\epsilon > 0$ (slightly underdamped) and along the imaginary direction for $\epsilon < 0$ (slightly overdamped).

To derive the above properties, consider the eigenvalue equation. For the moment, focus on one block; inter-block couplings will be added in Section~\ref{sect:jbperthigh} without difficulty. From (\ref{eq:jbh}), for its lowest-order determinant, $\H - \om$ is effectively represented by
\beq
  \bar{\H}^{\bdot}{}_{\bdot}-\om
  =\mbox{\small$
  \begin{pmatrix}\omega_j {-} \om & 1 & \cdots & 0 & 0 \\
     0 & \omega_j {-} \om & \ddots & 0 & 0 \\
     \vdots & \vdots & \ddots &  \ddots & \vdots \\
     0 & 0 & \cdots & \vphantom{\ddots}\omega_j {-} \om & 1 \\
     \epsilon \xi & 0 & \cdots & \vphantom{\ddots}0 & \omega_j  {-} \om
  \end{pmatrix}$}\;,
\eeql{eq:jbh1}
where in the determinant 
\beq
  \xi\equiv\langle \f^{j,M_j{-1}} | \Delta \H \f_{j,0} \rangle
  =(\f_{j,0},\Delta\H\f_{j,0})
\eeql{eq:jbp04}
multiplies the $1$'s, and all other elements of $\Delta \H$ are negligible because they
multiply another small quantity of at least order $\om_j-\om\equiv-\Delta\om$. Consequently, in
\beq
  J(\om) = (-1)^{M_j}[ J_0(\om) + \epsilon J_1(\om) + \epsilon^2 J_2(\om)+\cdots]\;,
\eeql{eq:pertchar1}
one has $J_0(\om) = (\om - \om_j)^{M_j}$ and $J_1(\om_j)=-\xi$. Setting $J(\om)=(-\Delta \om)^{M_j} + (-1)^{M_j{-}1}  \epsilon \xi = 0$, one finds
\begin{gather}
  \Delta \om_k = \lambda \zeta_k\;,\label{eq:jbp06}\\
  \lambda=(\epsilon \xi)^{1/M_j}\;,\qquad
  \zeta_k=e^{2i\pi k/M_j}\;;\label{eq:jbp07}
\end{gather}
$\lambda$ is any fixed choice of the root measuring the magnitude of the shift, $\zeta_k$ displays the equiangular behavior, and $k = 0, \ldots, M_j{-}1$. A change in $\sgn\epsilon$ costs a phase $e^{i\pi/M_j}$, causing the directions to bisect the original ones.

Incidentally, for the conservative case with degeneracies, (\ref{eq:jbh1}) would not have the $1$'s and the crucial term $\xi$ does not appear at this low order.

The eigenvectors can be expressed as~\cite{note1}
\beq
  \f_k = \sum_{n=0}^{M_j{-}1} \f_{j,n} \, T^n{}_{k}\;;
\eeql{eq:splitbas1}
upon setting $\om = \om_k$ in (\ref{eq:jbh1}), one solution is
\beq
  T^n{}_k = (\lambda \zeta_k)^n
\eeql{eq:p01}
[so $(T^{-1})^k{}_n =(\lambda \zeta_k)^{-n}\!/M_j$], with normalization
\beq
  (\f_k,\f_k)=M_j(\lambda\zeta_k)^{M_j-1}\;.
\eeql{eq:normsplit}

The above applies to any $\H$ cast into Jordan normal form.  But in the present case, $\H$ is given by (I.2.3); in particular, we assume that $\Delta\H$ does not affect the coupling to the bath, so $\Delta \Gamma = 0$ as in (I.4.1). Then,
\beq
  \xi = f_j(\alpha) \Delta K(\alpha,\beta) f_j(\beta)=(\Delta K)_{jj}\;,
\eeql{eq:jbp08}
in terms of the coordinates only. In the example of I, Section III A, if we let $K = k_* + \epsilon$, i.e., $\Delta K = 1$, then (\ref{eq:jbp08}) gives $\xi=1$ so (\ref{eq:jbp06}) gives $\Delta\om_k=\sqrt{\epsilon}e^{i\pi k}$, as expected.

\subsection{Higher orders}
\label{sect:jbperthigh}

To deal with higher-order corrections, we rewrite
\beq
  \H = \H_0 + \epsilon \Delta \H = \H_0' + \epsilon \Delta \H'\;,
\eeql{eq:jbp21}
shifting the term $\epsilon \xi$ to the unperturbed part, so that (\ref{eq:splitbas1}) diagonalizes $\H_0'$. In other words, $\langle\f^{j,M_j-1}|\Delta\H'\f_{j,0}\rangle=0$.

The matrix elements of $\Delta \H'$ in the split basis (\ref{eq:splitbas1}) follow by transforming those in the Jordan normal basis,
\beq\begin{split}
  (\Delta \H')^{k}{}_{k'}&=
  (T^{-1})^k{}_n \, \langle \f^{j,n} | \Delta \H' \f_{j,n'} \rangle
   \, T^{n'}{}_{k'} \\
  &= \frac{1}{M_j} \sum_{n,n'=0}^{M_j-1}\zeta_{k'}^{n'}\zeta_k^{-n}\lambda^{n'-n}
  \langle \f^{j,n} | \Delta \H' \f_{j,n'} \rangle\;.
\end{split}\eeql{eq:jbp23}
Because the term ($n'{=}0$, $n{=}M_j{-}1$) has been removed to~$\H_0'$, the lowest power of $\lambda$ is $\lambda^{2{-}M_j}$.  Together with $\epsilon = \lambda^{M_j}$, the leading correction is $O(\lambda^2)$, one power of $\lambda$ higher than the effect due to $\xi$ treated in Section~\ref{sect:jbpertlow}; it can be handled by the standard perturbation theory as in~I. Each higher-order correction will involve an extra matrix element $\propto\lambda^2$, divided by $\om_k - \om_{k'} \propto \lambda$, giving one overall power of $\lambda$ per order. Inter-block matrix elements of $\Delta \H'$ can be handled in the split basis as well.

In fact, there is not much to be gained by further analytic treatment; one can simply diagonalize the relevant $M_j \times M_j$ block of $\H$ numerically and deal with inter-block interactions in the resultant basis, using the non-degenerate perturbation theory of~I.

\subsection{Non-generic perturbations}
\label{sect:jbpertnongen}

The perturbation of a JB is said to be non-generic when $\xi = (\Delta K)_{jj} = 0$.  The simplest way to proceed is to expand (\ref{eq:pertchar1}) around
$\om_j$,
\beq\begin{split}
  0 &= (-1)^{M_j} J(\om) \\
  &= (\Delta \om)^{M_j}+\epsilon[J_1(\om_j)+\Delta\om J'_1(\om_j)+\cdots\,] \\
  &\quad + \epsilon^2 [ J_2(\om_j) + \cdots \, ] + \cdots\;.
\end{split}\eeql{eq:ngpert01}
In this case, $J_1(\om_j) = 0$; assuming there is no higher-order nongenericity, i.e., $J'_1(\om_j) \neq 0$, (\ref{eq:ngpert01}) gives
\beq
  0 = \Delta \om [ (\Delta \om)^{M_j-1} + \epsilon J'_1(\om_j)+ \cdots ]+ \cdots\;.
\eeql{eq:ngpert03}
Thus we see that (a)~one state is unshifted to lowest order, and (b)~the other states split like a generic JB of order $M_j-1$, viz., with $|\Delta\om| \propto \epsilon^{1/(M_j-1)}$ and splitting in $M_j-1$ equiangular directions. [If $M_j=2$, the $\epsilon^2 J_2(\om_j)$ term is of the same order and must be retained.] This situation will be seen in some of the examples in Section~\ref{sect:ex}. The exercise analogous to (\ref{eq:jbh1})--(\ref{eq:jbp04}) and (\ref{eq:jbp08}) shows that $J'_1(\om_j)=-2\xi'$, with
\beq
  \xi'=(\f_{j,1},\Delta\H\f_{j,0})
  =f_{j,1}(\alpha) \Delta K(\alpha,\beta) f_{j,0}(\beta)\;.
\eeql{xipr}

Not much is gained by formally pursuing nongeneric perturbations any further, since in applications it is again preferable to simply diagonalize the (typically small) JB.

\subsection{Small denominators near criticality}
\label{subsect:small}

The above resolves the ``small-denominator problem". Time evolution \emph{near} a critical point $\om_j$ is given by
\beq
  \bm{\phi}(t)={\sum_{k}}'e^{-i\om_k t}\f_k
  \frac{(\f_k,\bm{\phi})}{(\f_k,\f_k)}\;,
\eeql{eq:jbinm01}
where the prime denotes restriction to one \emph{cluster} of modes $\f_k$, which are close to merging at $\om_j$ as parameters are tuned. Already in part~I it has been argued that then $(\f_k,\f_k)\rightarrow0$ while $(\f_k,\bm{\phi})$ is finite, apparently causing $\bm{\phi}(t)$ to diverge. However, by using (\ref{eq:splitbas1})--(\ref{eq:normsplit}) and also (\ref{eq:jbp06}) as $e^{-i\om_k t} = e^{-i\om_j t} \sum_{l\ge0}(-i \lambda \zeta_kt)^l\!/l!
  = \sum_{l\ge0} (\lambda \zeta_k)^l C_l(\om_j,t)$, we get
\beq
  \bm{\phi}(t)=\sum_{l\ge0}\frac{C_l(\om_j,t)}{M_j}\sum_{n,n'=0}^{M_j-1}
  \f_{j,n}(\f_{j,n'},\bm{\phi})\sum_k(\lambda\zeta_k)^m\;,
\eeql{eq:jbinm03}
where $m=n+n'+l+1-M_j$ can be negative. But by (\ref{eq:jbp07}), $\sum_k$ vanishes unless $m\equiv0\pmod{M_j}$. Since the above is only valid to leading order anyway, we only keep $m=0$ terms, so $\sum_k\cdots=M_j$ and (\ref{eq:jbinm03}) reduces to $\bm{\phi}(t)=\sum_n\f_{j,n}(t)\langle\f^{j,n}|\bm{\phi}\rangle$ by (\ref{eq:time06}). Thus, the small denominators cancel (i.e., negative powers of $\lambda$ do not appear): while the contributions \emph{per mode} to $\bm{\phi}(t)$ are large, their \emph{sum} remains finite. Moreover, the latter agrees with (\ref{eq:jbgr07}), obtained in the Jordan normal basis \emph{at}~$\om_j$ [to lowest order; in general, $\bm{\phi}(t)$ smoothly depends on system parameters as these go through their critical values].

Similar arguments prove the cancellation of small denominators in other physical quantities, or for other splitting patterns (Section~\ref{sect:jbpertnongen}). These exactly parallel the discussion of excess noise (over the standard Schawlow--Townes value) in lossy laser cavities~\cite{excess,Pet}. In fact, only the noise per mode is enhanced while the sum over all (non-orthogonal) modes has no excess contribution, consistent with the fluctuation--dissipation theorem. Only perturbation theory requires special care: while a near-critical mode cluster shifts other eigenvalues by a finite amount, the cluster's modes themselves are highly sensitive to perturbations (Section~\ref{sect:jbperthigh})---indeed, no cancellation is expected in the properties of \emph{individual} modes.

%==========================================================

\section{Examples of higher-order criticality}
\label{sect:ex}

We construct some examples with JBs of size $M \ge 2$.  It suffices to consider an $N=2$ system, with 
\beq
  K = \begin{pmatrix} k_{11} & k_{12} \\ k_{12} & k_{22} \end{pmatrix}\;, \quad
  \Gamma = 2 \, \begin{pmatrix} \gamma_{11} & \gamma_{12} \\
    \gamma_{12} & \gamma_{22} \end{pmatrix}\;,
\eeql{eq:jbex01}
involving 6 parameters in general. 

\subsection{Fourth-order JB}
\label{subsect:ex1}

To find an $M=4$ JB, let $J(\om)$ have a 4th-order zero:
\beq
  \det( \H^{\bdot}{}_{\bdot} - \om )= ( \om + ia )^4
\eeql{eq:jbex02}
(necessary but not sufficient, since equality of eigen\emph{values} could also indicate level crossing; the further conditions will be considered later). The position of the root has to be on the negative imaginary axis (or else there would be another root at $-ia^*$).  Without loss of generality, we henceforth set $a=1$, and the $a$-dependence can eventually be restored by scaling $K \mapsto a^2 K$, $\Gamma \mapsto a \Gamma$,
$\om \mapsto a \om$.

Eliminating the momenta in $(\H^{\bdot}{}_{\bdot} - \om)\bm{\phi} = 0$, i.e., setting $\om = -i$ in (I.2.7), gives
\beq
  (K - \Gamma + I ) \,\phi = 0\;.
\eeql{eq:jbex09}
Generically, there is only one independent $\phi$ satisfying (\ref{eq:jbex09}), hence only one eigenvector $\bm{\phi}$, hence an $M=4$ JB.

Equation (\ref{eq:jbex02}) leads to four conditions for the coefficients of $\om^3, \ldots, \om^0$, which after some simplification give
\beq\begin{split}
  k_{11} k_{22} - k_{12}^2 &= 1\\
  \gamma_{11} + \gamma_{22} &= 2\\
  k_{11} + k_{22} + 4( \gamma_{11} \gamma_{22} - \gamma_{12}^2 ) &= 6\\
  k_{11} \gamma_{22} + k_{22} \gamma_{11} - 2 k_{12}\gamma_{12} &= 2\;.
\end{split}\eeql{eq:jbex04}
The general solution for $K$ involves two parameters:
\beq
  K=\begin{pmatrix} e^y \cosh x & \sinh x \\ \sinh x & e^{-y} \cosh x \end{pmatrix}\;,
\eeql{eq:jbex05}
solving the first of the four constraints. For any $x,y$, the three remaining equations determine the three~$\gamma_{ij}$. Positivity of $K$ is guaranteed, while $\Gamma\ge0$ requires $\cosh x\cosh y\le 3$. A simple choice (which satisfies this as an equality) is $\sinh x=-2$, $e^{2y}=5$, giving
\beq
  K=\begin{pmatrix}\phantom{-}5 & -2 \\ -2 & \phantom{-}1\end{pmatrix}\;,\qquad
  \Gamma=\begin{pmatrix} 4 & 0 \\ 0 & 0 \end{pmatrix}\;.
\eeql{eq:jbex09b}

The Jordan normal basis is constructed to be
\beq\begin{split}
  \f_{,0} &= \sqrt{2}i\,(1,1,-1,-1)^{\rm T}\\
  \f_{,1} &= {\ts\frac{1}{2}}\sqrt{2}\,(-1,1,3,1)^{\rm T}\\
  \f_{,2} &= {\ts\frac{1}{8}}\sqrt{2}i\,(-1,-1,5,-3)^{\rm T}\\
  \f_{,3} &= {\ts\frac{1}{16}}\sqrt{2}\,(-1,1,-1,-3)^{\rm T}\;,
\end{split}\eeql{eq:jbex10}
normalized as in (\ref{eq:summ1}). These vectors are arbitrary up to \emph{one} overall sign, and exemplify the formalism in Section~\ref{sect:jb}. In particular, the alternation of real and imaginary basis vectors is prescribed by (\ref{JBconj}), here realized with the lower sign. The duals follow from (\ref{eq:jbdual}) as
\beq\begin{split}
  \f^{,0} &= {\ts\frac{1}{16}}\sqrt{2}i\,(5,3,1,-1)^{\rm T}\\
  \f^{,1} &= {\ts\frac{1}{8}}\sqrt{2}\,(-1,3,1,1)^{\rm T}\\
  \f^{,2} &= {\ts\frac{1}{2}}\sqrt{2}i\,(1,-1,1,-1)^{\rm T}\\
  \f^{,3} &= \sqrt{2}\,(-3,1,-1,-1)^{\rm T}\;,
\end{split}\eeql{eq:jbex10a}
and are readily verified to obey (\ref{eq:orthjb}). 

It is possible to have more than one $\phi$ satisfy (\ref{eq:jbex09}). For a $2 \times 2$ system, this requires $K-\Gamma+I =\nobreak 0$, which together with (\ref{eq:jbex05}) fixes $\gamma_{ij}$.  When put back into the three remaining equations in (\ref{eq:jbex04}), these lead to one additional condition $\cosh x\cosh y=1$, which has only the solution $x = y =0$---the trivial case of two independent but identical oscillators, each generating an $M=2$ JB at the critical point. Apparently, with $N=2$, one cannot have a crossing between $M=3$ and $M=1$ blocks. Thus, level crossing occurs only exceptionally, under the additional condition that a minor determinant of $\H^{\bdot}{}_{\bdot}-\om$ vanishes. The JBs of the two oscillators can be mixed, illustrating the subtleties encountered in Section~\ref{subsect:degen}. However, taking complex superpositions will in general break the symmetry (\ref{JBconj}), generalizing the discussion below (I.2.14).

\begin{figure}
 \includegraphics[width=3in]{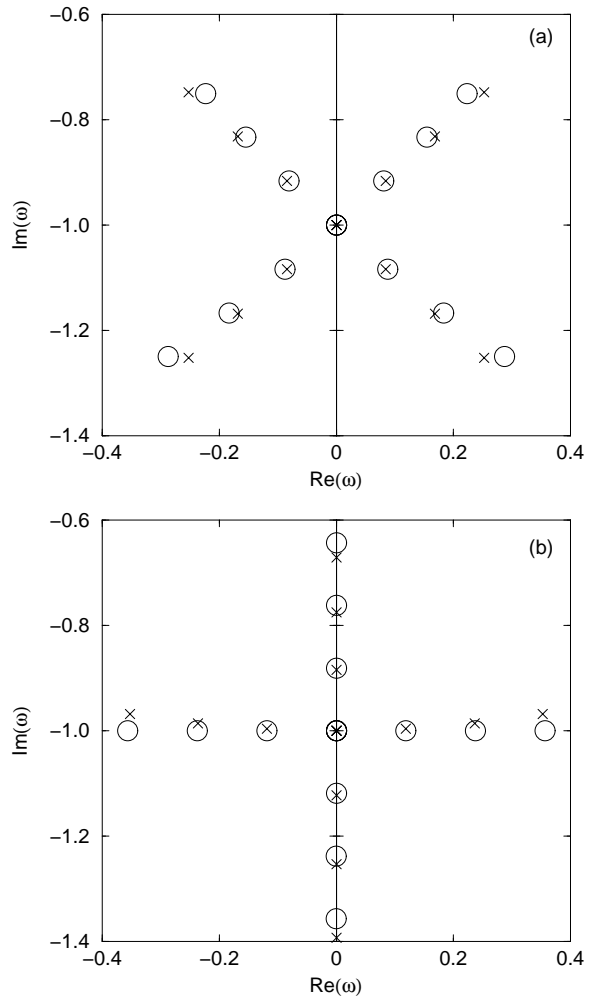}
 \caption{Eigenvalues for an
$M = 4$ JB split by
$k_{11} \mapsto k_{11} + \epsilon$, with
$\epsilon = n^4 \epsilon_0$, 
$n = 0, 1, 2, \ldots$.  
(a)~$\epsilon_0 = 10^{-4}$;
(b)~$\epsilon_0 = -10^{-4}$.
Crosses (circles) denote numerical (perturbative) values.
The nearly equal spacing shows that the shifts
are $\propto\epsilon^{1/4}$.}\label{fig1}
\end{figure}

\begin{figure}
 \includegraphics[width=3in]{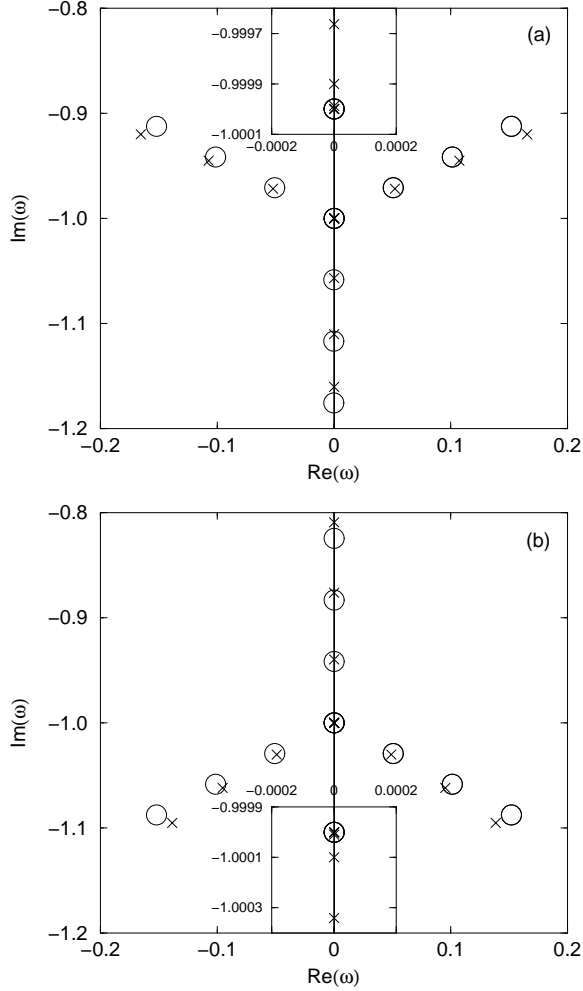}
\caption{Eigenvalues for an
$M = 4$ JB split by the non-generic perturbation
$k_{ij} \mapsto k_{ij} + \epsilon \mu_{ij}$,
$\mu_{11}=\nobreak1$, $\mu_{12}=-\frac{3}{2}$, $\mu_{22} = 2$;
$\epsilon = n^3 \epsilon_0$, 
$n = 0, 1, 2, \ldots$.  
(a)~$\epsilon_0 = 10^{-4}$;
(b)~$\epsilon_0 = -10^{-4}$.
Crosses (circles) denote numerical (perturbative) values.
Three of the eigenvalues split at $120^\circ$; the
nearly equal spacing shows that the shifts are
$\propto\epsilon^{1/3}$.  The fourth, nearly
unchanged, eigenvalue is shown in greater detail in the inset.}\label{fig2}
\end{figure}

It is instructive to consider perturbations around the $M=4$ JB
(\ref{eq:jbex09b})--(\ref{eq:jbex10a}).  First, let $k_{11} \mapsto k_{11} + \epsilon$, and evaluate the lowest-order result (\ref{eq:jbp06}) (with $\xi = -2$) for various~$\epsilon$; these are shown by the circles in Fig.~\ref{fig1}, illustrating the properties below (\ref{eq:jbp01}). The exact numerical eigenvalues, shown by the crosses in Fig.~\ref{fig1}, demonstrate the accuracy of perturbation theory. Second, consider the more general $k_{ij} \mapsto k_{ij} + \epsilon \mu_{ij}$, with $\mu_{ij}=O(1)$. By expanding $J(\om)$ for the model (\ref{eq:jbex01}) as in (\ref{eq:pertchar1}), we find
\beq\begin{split}
  J_1(-i) &= (k_{22}{+}1{-}2\gamma_{22} ) \mu_{11}
  + (k_{11}{+}1{-}2\gamma_{11}) \mu_{22}\\
  &\quad+2(2\gamma_{12}{-}k_{12}) \mu_{12}\;.
\end{split}\eeql{eq:jbex12}
For the parameters of (\ref{eq:jbex09b}), this vanishes if
$\mu_{11} = 1$,
$\mu_{12} = \nobreak-\frac{3}{2}$,
$\mu_{22} = 2$,
yielding a non-generic perturbation [$\xi=0$ in (\ref{eq:jbp08})]. Figure~\ref{fig2} shows the eigenvalues vs $\epsilon$ (crosses and circles as in Fig.~\ref{fig1}). To leading order one is not shifted, while the others split like a generically perturbed $M=3$ block (i.e., $\sim\epsilon^{1/3}$, smaller than the typical shift $\sim\epsilon^{1/4}$)---in quantitative agreement with the perturbative (\ref{eq:ngpert03}) and (\ref{xipr}), where presently $\xi'=i$.

\subsection{Third-order JB}
\label{subsect:ex2}

Next consider a third-order JB, mainly to show that odd $M \neq 1$ is allowed. For $N=2$, this is achieved by setting $J(\om)= ( \om + i )^3 (\om + ib)$, with $b \neq 1$. The triple root has been scaled to $-i$. As before, one obtains
\beq\begin{split}
  k_{11} k_{22} - k_{12}^2 &= b\\
  \gamma_{11} + \gamma_{22} &= (3 + b)/2\\
  k_{11}+k_{22}+4(\gamma_{11}\gamma_{22}{-}\gamma_{12}^2)&=3(1+b)\\
  k_{11}\gamma_{22}+k_{22}\gamma_{11}-2k_{12}\gamma_{12}&=(1+3b)/2\;.
\end{split}\eeql{eq:jbex24}
We work in the eigenbasis of $\Gamma$ (cf.\ Section III~B and note [3], both in~I) from the outset, and can solve for the remaining parameters in terms of $\gamma_{11}$ and~$b$. This involves rationals only for, e.g., $b=4$ and $K=\frac{1}{5}\bigl(\begin{smallmatrix}41 & 8 \\ 8 & 4\end{smallmatrix}\bigr)$, $\Gamma=\bigl(\begin{smallmatrix} 6 & 0 \\ 0 & 1 \end{smallmatrix}\bigr)$, where for variation we took a $\Gamma>0$.

The basis vectors are found to be
\beq\begin{split}
  \f_{1,0} &= e^{i\pi/4}\sfrac{\sqrt{15}}{15}\,
    (2,-4,-2,4)^{\rm T}\\
  \f_{1,1} &= e^{-i\pi/4}\sfrac{\sqrt{15}}{180}\,
    (-19,-22,43,-26)^{\rm T}\\
  \f_{1,2} &= e^{i\pi/4}\sfrac{\sqrt{15}}{2880}\,
    (-221,-78,525,430)^{\rm T}\\
  \f_{2,0} &= e^{i\pi/4}\sfrac{\sqrt{15}}{45}\,
    (8,-1,-32,4)^{\rm T}\;.
\end{split}\eeql{eq:jbex26}
It is seen that the $j=1$ ($j=2$) block obeys (\ref{JBconj}) with the lower (upper) sign; apparently, not much can be said about this sign in general. The dual vectors are 
\beq\begin{split}
  \f^{1,0} &= e^{i\pi/4}\sfrac{\sqrt{15}}{2880}\,
    (801,-352,221,78)^{\rm T}\\
  \f^{1,1} &= e^{-i\pi/4}\sfrac{\sqrt{15}}{180}\,
    (-71,-48,-19,-22)^{\rm T}\\
  \f^{1,2} &= e^{i\pi/4}\sfrac{\sqrt{15}}{15}\,
    (-10,0,-2,4)^{\rm T}\\
  \f^{2,0} &= e^{i\pi/4}\sfrac{\sqrt{15}}{45}\,
    (-16,-3,-8,1)^{\rm T}\; ,
\end{split}\eeql{eq:jbex26a}
again up to an overall sign.

\begin{figure}
 \includegraphics[width=3in]{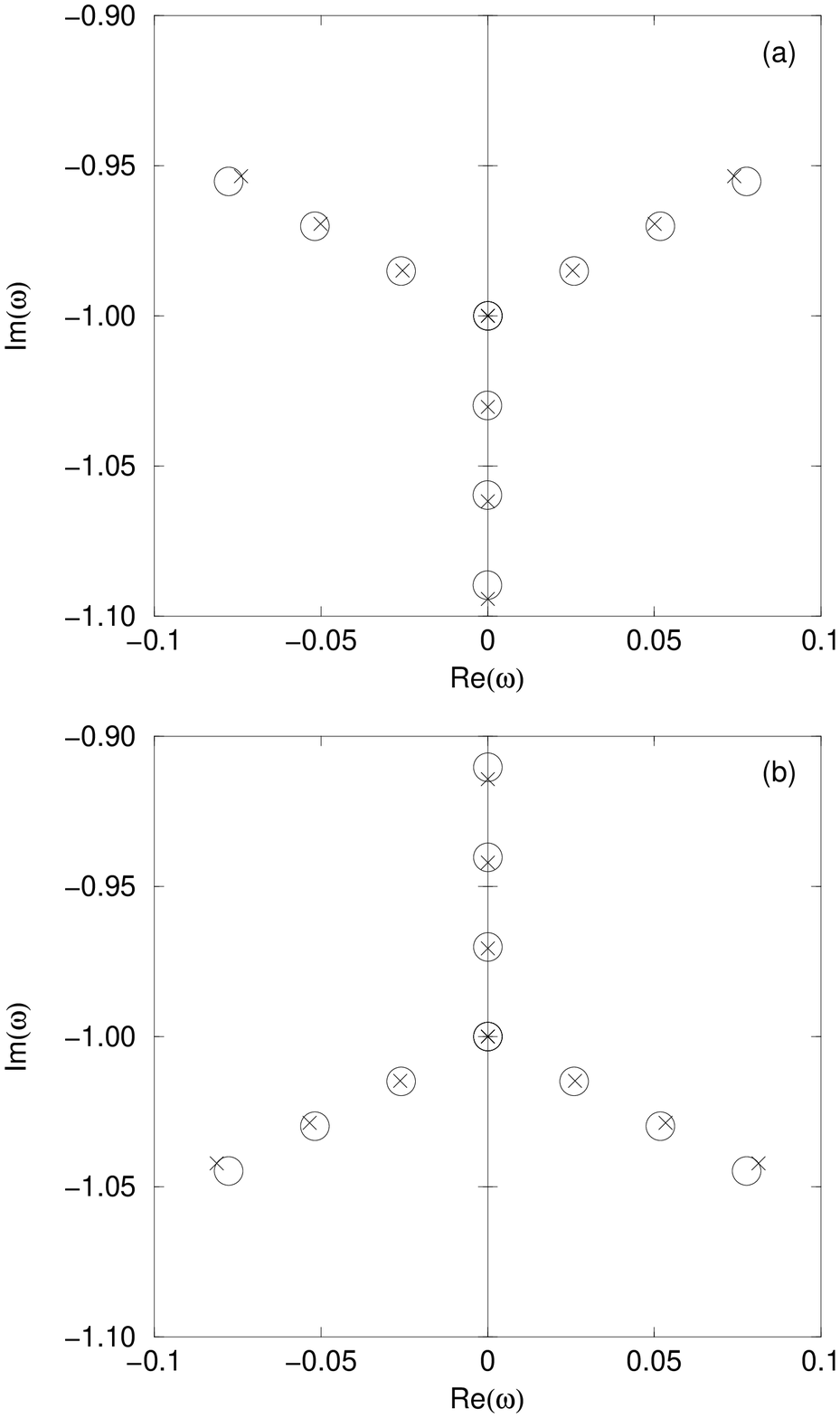}
\caption{Eigenvalues for an
$M = 3$ JB split by
$k_{11} \mapsto k_{11} + \epsilon$, with
$\epsilon = n^3 \epsilon_0$, 
$n = 0, 1, 2, \ldots$.  
(a)~$\epsilon_0 = 10^{-4}$;
(b)~$\epsilon_0 = -10^{-4}$.
Crosses (circles) denote numerical (perturbative) values.
The nearly equal spacing shows that the shifts are $\propto\epsilon^{1/3}$.}
\label{fig3}
\end{figure}

\begin{figure}
 \includegraphics[width=3in]{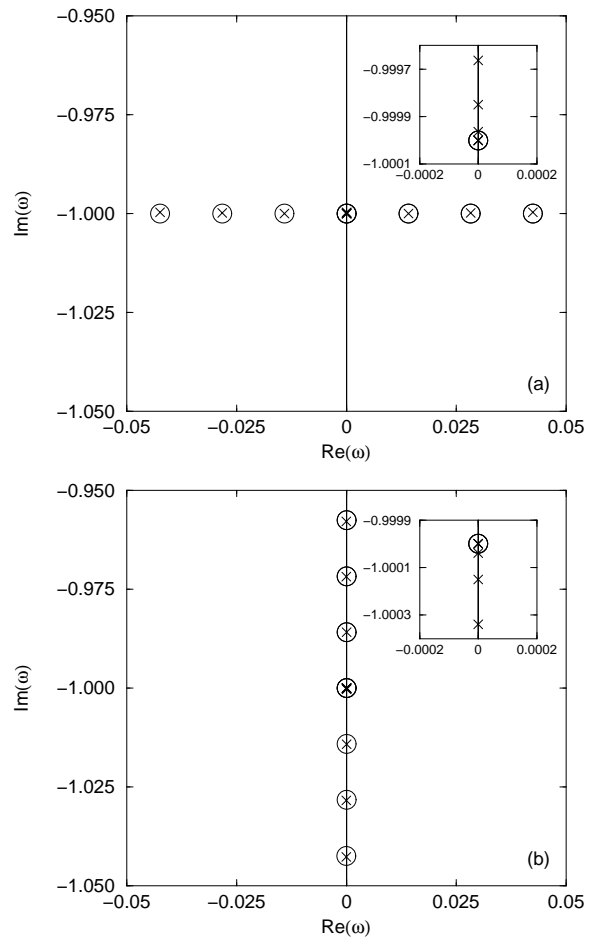}
\caption{Eigenvalues for an
$M = 3$ JB split by
$k_{ij} \mapsto k_{ij} +\nobreak\epsilon \mu_{ij}$, with
$\mu_{11}=-2$, $\mu_{22} = 1$, $\mu_{12}=1/2$ chosen 
so that the perturbation is non-generic, and
$\epsilon = n^2 \epsilon_0$, 
$n = 0, 1, 2, \ldots$.  
(a)~$\epsilon_0 = 10^{-4}$;
(b)~$\epsilon_0 = -10^{-4}$.
Crosses (circles) denote numerical (perturbative) values.
Two of the eigenvalues split at $180^\circ$; the
nearly equal spacing shows that the shifts are
$\propto\epsilon^{1/2}$. The third eigenvalue is nearly
unchanged, and is shown in greater detail in the inset.}\label{fig4}
\end{figure}

Next change $k_{11} \mapsto k_{11} + \epsilon$, giving $\xi = \frac{4}{15}i$; Fig.~\ref{fig3} shows the eigenvalues emanating from the JB (conventions as before). For $k_{ij} \mapsto k_{ij} +\epsilon \mu_{ij}$, the choice $\mu_{11} = -2$, $\mu_{12} = \frac{1}{2}$, $\mu_{22} = 1$ makes $\xi=0$, eliminating the leading term, while $\xi'=1$ in the next order. Figure~\ref{fig4} shows the eigenvalues; to leading order, one is not shifted, while the other two split like a generically perturbed $M=2$ block.

\subsection{Two second-order JBs}
\label{subsect:ex3}

So far, the examples only involve JBs with imaginary frequencies. This is not necessary: consider two second-order JBs at $\om = -i \pm b$, by setting $J(\om) = (\om + i - b)^2 \* (\om + i + b)^2$. As before, this leads to
\beq\begin{split}
  k_{11} k_{22} - k_{12}^2 &= (1+b^2)^2\\
  \gamma_{11} + \gamma_{22} &= 2\\
  k_{11}+k_{22}+4(\gamma_{11}\gamma_{22}-\gamma_{12}^2)&=6+2b^2\\
  k_{11}\gamma_{22}+k_{22}\gamma_{11}-2k_{12}\gamma_{12}&=2(1+b^2)\;.
\end{split}\eeql{eq:jbex32}
Guided by Section~\ref{subsect:ex1}, without loss of generality we proceed in the eigenbasis of $\Gamma$. With $\gamma_{12}=0$, (\ref{eq:jbex32}) can be readily solved for the remaining parameters in terms of $b$ and~$\gamma_{11}$. Again we consider the marginal case $\gamma_{11}=2$, and a final simplification occurs for $b=\frac{4}{3}$, in which case all $k_{ij}$ turn out rational: $K=\frac{1}{9}\bigl(\begin{smallmatrix} \phantom{-}61 & -30 \\ -30 & \phantom{-}25\end{smallmatrix}\bigr)$ and $\Gamma=\bigl(\begin{smallmatrix} 4 & 0 \\ 0 & 0 \end{smallmatrix}\bigr)$.

The basis vectors read
\beq\begin{split}
  \f_{1,0}&=\sfrac{\sqrt{6}}{24}\left(3{-}6i,-3{-}6i,
    -11{+}2i,-5{+}10i\right)^{\rm T}\\
  \f_{1,1}&=\sfrac{\sqrt{6}}{192}\left(15{+}30i,
    -15{+}30i,-23{-}74i,7{+}14i\right)^{\rm T},
\end{split}\eeql{eq:jbex34}
the conjugates $\f_{-1,n}$ following from (\ref{JBconj}). Their duals,
\begin{align}
  \f^{1,0}&=\sfrac{\sqrt{6}}{192}\left(-46{-}37i,-14{-}7i,
    -30{-}15i,-30{+}15i\right)^{\rm T}\notag\\
  \f^{1,1}&=\sfrac{\sqrt{6}}{24}\left(22{-}i,
    -10{+}5i,6{-}3i,6{+}3i\right)^{\rm T}\;,\label{eq:jbex35}
\end{align}
obey (\ref{eq:orthjb}) not only with the vectors in (\ref{eq:jbex34}), but also with their conjugates $\f_{-1,n}$.

\begin{figure}
 \includegraphics[width=3in]{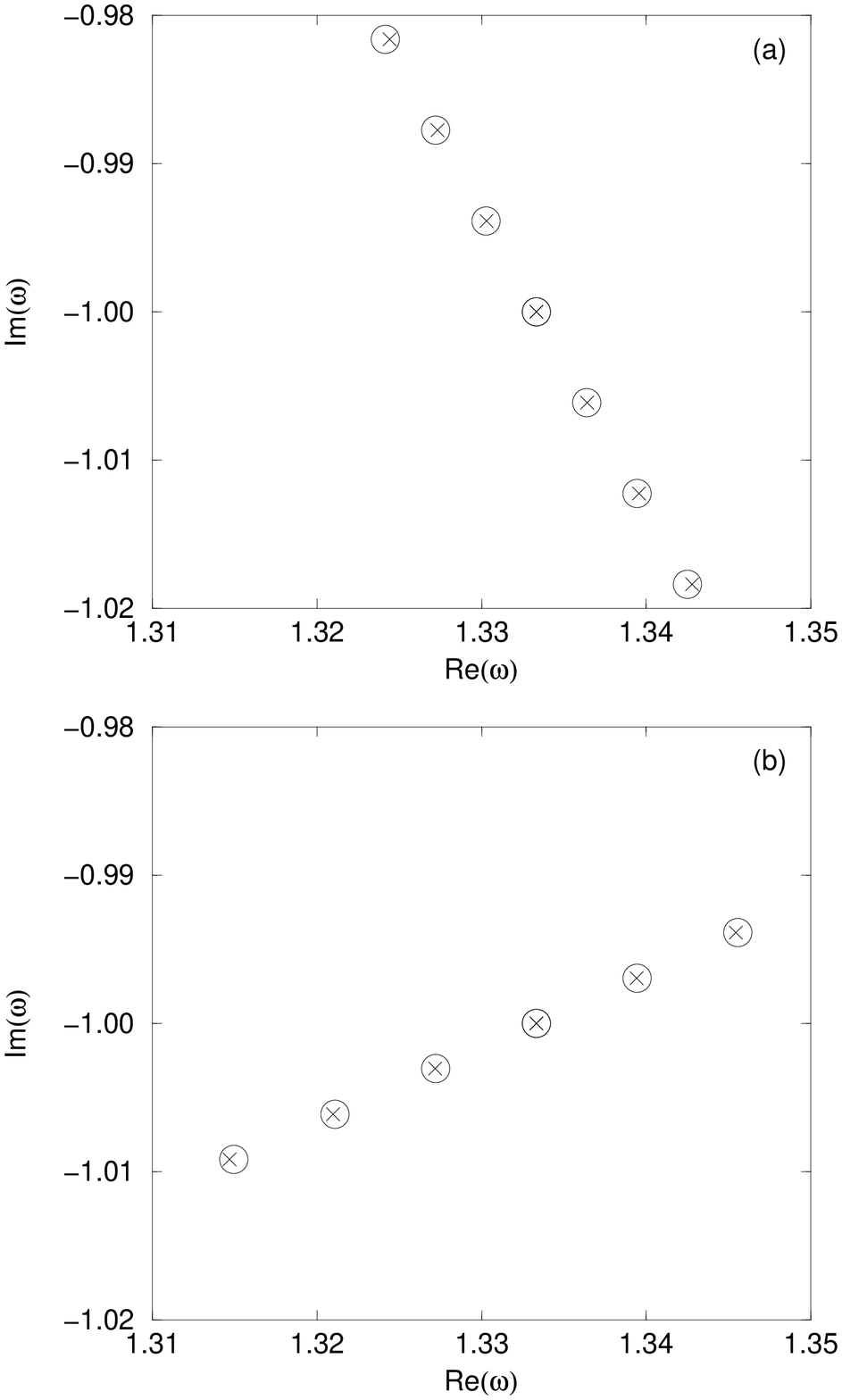}
\caption{Eigenvalues for two
$M = 2$ JBs split by
$k_{11} \mapsto k_{11} + \epsilon$, with
$\epsilon = n^2 \epsilon_0$, 
$n = 0, 1, 2, \ldots$.  
(a)~$\epsilon_0 = 10^{-4}$;
(b)~$\epsilon_0 = -10^{-4}$.
Crosses (circles) denote numerical (perturbative) values.
Only the right-half $\om$-plane is shown.
The nearly equal spacing shows that the shifts are
$\propto\epsilon^{1/2}$.}\label{fig5}
\end{figure}

Figure~\ref{fig5} shows the eigenvalues when $k_{11} \mapsto k_{11} + \epsilon$, where in this case $\xi = -(9{+}12i)/32$. (Because of symmetry, only the right-half $\om$-plane is shown.)

Finally, consider (\ref{eq:jbex32}) for variable~$b$, keeping $\Gamma=\diag(4,0)$ fixed. One readily solves for $K(b)$, which [choosing $k_{12}(b)<0$] for $b\rightarrow0$ tends to $K$ as in (\ref{eq:jbex09b}). Reversing the procedure, one thus has found a highly nongeneric perturbation splitting an $M=4$ JB into two $M=2$ JBs, which do not undergo further splitting. Note that the family $\H(b)$ is \emph{not} of the form (\ref{eq:jbp01}): it is impossible to obtain this particular structure with only a first-order correction $\Delta\H$, if the latter corresponds to $\Delta K\neq0$ only. See Sections 6.2 and~6.3 (final paragraph) in Ref.~\cite{jbsr}, where the existence of off-axis JBs in a continuum model is studied as an open question.

%==========================================================

\section{Discussion}
\label{sect:disc}

We have extended the eigenvector expansion developed in the previous paper~\cite{pap1} to situations where the eigenvectors merge and thus are incomplete. The Jordan normal basis is then used. Such a basis for a general matrix operator $\H$ (not self-adjoint) is well known, but here its properties have to be considered together with the bilinear map.  The bi-orthogonality of the Jordan normal basis, most simply expressed by the metric (\ref{eq:jbflip}), has been established: part of it being intrinsic and part of it being a conventional choice. As a consequence, $\H_{\bdot\bdot}$ takes the \emph{symmetric} Jordan-type form~(\ref{eq:jbh2}).

The Jordan basis vectors (like the eigenvectors in the non-critical case) immediately solve the dynamics. The time evolution of these vectors $\f_{j,n}$ is characterized by a polynomial prefactor in~$t$, as expected when several exponential terms with slightly different frequencies merge.  

Perturbations around critical points are particularly interesting, a term $\epsilon \Delta \H$ shifting frequencies by a \emph{fractional} power of $\epsilon$ in equiangular directions. Their study allows the small-denominator problem
associated with $(\f_k , \f_k) \rightarrow 0$ at criticality to be handled.

All these concepts have been illustrated by nontrivial examples. By treating criticality, this paper complements the previous one~\cite{pap1}, and places the familiar concept of critical damping into a general framework. As in~\cite{pap1}, the entire formalism can be promoted to the quantum domain by turning the state vectors into operators.

%==========================================================

\acknowledgments

This work is built upon a long collaboration with E.S.C. Ching, H.M. Lai, P.T. Leung, S.Y. Liu, W.M. Suen, C.P. Sun, S.S. Tong and many other colleagues.  KY thanks R.K. Chang for discussions on microdroplet optics, initiating our interest in waves in open systems. We thank S.L. Cheung for help with the perturbative examples. AMB was supported by a C.N. Yang Fellowship.

%==========================================================


\begin{thebibliography}{99}
\bibitem{LleN}
A trivial constraint here is that the eigenvectors follow from the \emph{one}-component (I.2.7). Hence, $L\le N$, as will be exemplified in Section~\ref{subsect:ex1}.
\bibitem{note1}
The $k$-th eigenvalue to split off from (or merge into) a JB labeled as $j$ should be denoted as, e.g., $\om_{j(k)}$, $k = 0, \ldots, M_j{-}1$; to avoid overly cumbersome notation, we shall simply write~$\om_k$ (and likewise for the eigenvectors). This is also appropriate in that for a finite splitting, it is no longer possible to associate the $\om_k$ unambiguously with a block $j$ from which they split.
\end{thebibliography}
\end{document}